\documentclass[epj]{svjour}
\usepackage{graphics}
\usepackage{amsmath}
\usepackage{graphicx,epsfig,psfrag}
\usepackage{amssymb}

\newcommand{\OLD}[1]{{\bf OLD RM}}

\renewcommand{\vec}[1]{{\mathbf #1}}

\newcommand{\w}{\omega}
\newcommand{\e}{\epsilon}
\begin{document}
\title{Breakdown of Luttinger's theorem in two-orbital Mott insulators}
\author{A. Rosch
}                     
%
%
\institute{Institute for Theoretical Physics, University of Cologne, 50937
Cologne, Germany}
\date{Received: date / Revised version: date}
%
\abstract{  An analysis of Luttinger's theorem shows that -- contrary to recent
  claims -- it is {\em not} valid for a generic Mott insulator. For a
  two-orbital Hubbard model with two electrons per site the crossover
  from a non-magnetic correlated insulating phase (Mott or Kondo
  insulator) to a band insulator is investigated. Mott insulating
  phases are characterized by poles of the self-energy and
  corresponding zeros in the Greens functions defining a ``Luttinger
  surface'' which is absent for band insulators.  Nevertheless, the
  ground states of such insulators with two electrons per unit cell
  are adiabatically connected.
} 
\maketitle
The basic quantity which defines a metal at low temperatures is
the Fermi surface. Excitations of the Fermi surface are the basis
of Fermi liquid theory. The concept of the Fermi surface also
allows to distinguish different phases. Changes in the topology of
the Fermi surfaces (e.g. vanishing bands) are therefore always
associated with quantum phase transitions.

Does an analog quantity exist in the case of a Mott insulator
where strong interactions prohibit the formation of a Fermi
surface? In a number of recent papers 
\cite{dzyaloshinskii,tsvelik1,tsvelik2,yang,stanescu,berthod} it has been
argued that in such a situation the concept of a Fermi surface has
to be replaced by the so-called ``Luttinger surface''. While the
Fermi surface is defined by poles of the Greens function, the
Luttinger surface is obtained from the zeros of the Greens
function or, equivalently, from the poles of the self-energy.

This identification is motivated by Luttinger's theorem
\cite{luttinger}, which states that at zero temperature, $T=0$,
the total density of electrons $n$ can be obtained by summing up
all momenta where the real part of the Greens function evaluated
at the Fermi energy is positive. For a multiband system the
propagator is a matrix. Denoting the eigenvalues of this matrix by
$G_\alpha(\w,\vec{p})$ and using that $\text{Im}\,
G_\alpha(0,\vec{p})=0$, Luttinger's theorem (also called Luttinger's sum rule) takes the form
\begin{equation}\label{luttinger}
n = 2 \sum_\alpha \int_{G_\alpha(\w=0,\vec{p})>0} \frac{d^3 \vec{p}}{(2 \pi)^3}
\end{equation}
It claims that the total density of electrons is fixed by the volume
enclosed by a surface where the zero-frequency propagator
$G(0,\vec{p})$ changes sign (alternative versions of the theorem are
briefly discussed in Sec.~\ref{discussion}). As emphasized recently by
Dzyaloshinskii \cite{dzyaloshinskii} and by Essler and
Tsvelik \cite{tsvelik1}, there are two ways how such a sign change can
happen. At a Fermi surface, $E(\vec{p})=\mu$, there is a pole,
$G(\w,\vec{p}) \approx \frac{Z_\vec{p}}{\w-(E(\vec{p})-\mu)}$, where
$Z_\vec{p}$ and $E_\vec{p}$ are the weight and dispersion of the quasi
particle. Alternatively, the propagator can change its sign going
through zero instead of infinity, $G(\w=0,\vec{p})=0$.  The latter
condition defines the ``Luttinger surface'' which is realized in Mott
insulators \cite{dzyaloshinskii,tsvelik1,tsvelik2}.  This concept was
for example used by Yang, Rice and Zhang \cite{yang} to build up a
phenomenological theory of the pseudogap state. In a recent preprint,
Stanescu, Phillips and Choy \cite{stanescu} tried to explore the role
of Luttinger surfaces for (doped) Mott insulators. Recently, Ortloff,
Balzer and Potthoff \cite{potthoff07} investigated under what
conditions various (non-perturbative) approximation schemes lead to a
violation of the Luttinger theorem.

This motivates us to investigate the question whether the concept
of such a Luttinger surface is as general and robust as the Fermi
surface. Can
one conclude that two states with a different topology of
Luttinger surfaces have to be separated by a quantum phase
transition? In Ref.~\cite{tsvelik2}, Konik, Rice and
Tsvelik have for example constructed a doped spin liquid (by
coupling Mott-insulating Hubbard ladders) which is characterized
both by a Luttinger surface and by the Fermi surfaces of
small particle and hole
pockets. Here the question arises whether this seemingly exotic
state of matter is adiabatically connected to a weakly interacting
system where Luttinger surfaces are absent.

In the following we will first show that Luttinger's theorem
(\ref{luttinger}) is not valid for a generic Mott insulator using
an explicit counter example. We will then discuss which
assumptions underlying its proof may not be fulfilled and finally
investigate implications for the question whether various insulating
states are adiabatically connected.

\section{Model}
For definiteness, the following two-band Hubbard model is
considered
\begin{eqnarray}
H&=&H_{kin}+ \sum_i H^i_{\text{loc}} \label{Htot}\\
H_{kin}&=& \sum_{ij\alpha} t_{ij}^\alpha
c^\dagger_{\alpha i \sigma}c_{\alpha j \sigma}\\
H^i_{\text{loc}}&=&  V ( c^\dagger_{1i\sigma} c_{2i\sigma} +
c^\dagger_{2i\sigma} c^{\ }_{1i\sigma}) + J
\vec{S}_{1 i}\vec{S}_{2 i} \nonumber \\
&& +\sum_{\alpha=1}^2 \left[(\epsilon_\alpha - \mu)
c^\dagger_{\alpha i \sigma} c_{\alpha i \sigma}+ U_\alpha
n_{\alpha i \uparrow} n_{\alpha i \downarrow} \right]
\end{eqnarray}
where $\alpha=1,2$ is the orbital index, $\vec{S}_{\alpha i}$ the
spin at site $i$ in band $\alpha$ and $n_{\alpha i \sigma}$ the
number of spin $\sigma$ electrons at site $i$ in band $\alpha$.
The model (\ref{Htot}) and various limits of it (which include the
Anderson lattice or Kondo lattice, models for bilayer compounds or
multi-orbital systems) have been widely studied.

The main motivation why we are interested in the model
(\ref{Htot}) in the context of this paper (rather than e.g. the
one-band Hubbard model) is that it has a simple, almost trivial
limit which allows us to discuss the validity of Luttinger's
theorem for Mott insulators and crossovers between different types
of insulators.  We consider the limit where (i) the system is
half-filled as there are two electrons per unit cell, (ii) the
ground-state of $H_{\text{loc}}^i$ is a singlet separated by a gap
$\Delta$ from all excited states (see Appendix~A),
and (iii) the hopping is very small, $t \ll \Delta$.  The physics
of this limit can be fully understood using a straightforward
strong coupling expansion around the atomic limit, $H_{kin}=0$
(performed e.g. up to 11th order in Ref.~\cite{monien} for
a Kondo insulator). In contrast, the Mott insulating phase of the
single-band Hubbard models is more difficult to analyze as the
ground-state of $H_{\text{loc}}$ is spin-degenerate which usually
leads to some form of magnetism (and also the zero-temperature 
non-magnetic state obtained e.g. within dynamical mean field theory is
non-generic and responds in a singular way to tiny magnetic fields \cite{potthoff07}).
Within our model one can easily
study the crossover from $\epsilon_1=\epsilon_2$, $V=0$ and
$U_\alpha \gg t^\alpha$ where one has two coupled Mott insulators
with $\Delta=3 J/4$ and charge gap $U/2+3 J/4$ to a band insulator
with $U_\alpha=0$ and a large band gap $ \Delta =
\sqrt{\left(\frac{\epsilon_1-\epsilon_2}{2}\right)^2+V^2}$. Taking
a slightly different limiting procedure [$U_2=2 (\mu -\epsilon_2)
\to \infty, t^2_{ij}=0, U_1=0, V=0$ and $J>0$], the model
(\ref{Htot}) also describes a Kondo insulator, i.e. the insulating
phase of a half-filled Kondo-lattice model.

\section{Breakdown of Luttinger's theorem}
 \subsection{Local limit}\label{local}
 First,  two coupled Mott insulators are analyzed with $V=0,
U_1=U_2=U, \epsilon_1=\epsilon_2=-U/2,
 t^1_{ij}=t^2_{ij}=t_{ij}$,
 $J,U \gg t_{ij}$ and $\mu$ chosen such that the system is half
 filled, $n=2$. For $t_{ij}=0$, the energy (per site) of the singlet ground-state is
$-2 \mu-3 J/4$. The $T=0$ Greens function of each of the two bands
is given by
\begin{eqnarray}\label{gLoc}
G_{\text{loc}}(\w)=\frac{1}{2} \left( \frac{1}{\w+\mu-\tilde{U}/2}
+\frac{1}{\w+\mu+\tilde{U}/2}  \right)
\end{eqnarray}
with $\tilde{U}=U+3 J/2$. The corresponding local self-energy is given
by
\begin{eqnarray}\label{sigmaLoc}
\Sigma_{\text{loc}}(\w)= \frac{\tilde{U}}{2}+
\frac{(\tilde{U}/2)^2}{\w+\mu}.
\end{eqnarray}
This form of the self-energy is well known from the Hubbard I
approximation \cite{hubbard1} which, indeed, correctly describes
the zeroth order strong-coupling expansion (note, however, that we
are not using the Hubbard I approximation in the following but
instead a {\em controlled} strong-coupling expansion). To leading
order in $t_{ij}$, the self-energy stays local and the Greens
function is obtained as
\begin{eqnarray}\label{leading}
G(\w,\vec{p})=\frac{1}{\w-[\e_\alpha+
t_{\vec{p}}-\mu+\Sigma_{\text{loc}}(\w)]}
\end{eqnarray}
where the single-particle dispersion  $t_{\vec{p}}$ is the Fourier
transform  of $t_{ij}$. To this order, the spectral function
at momentum $\vec{p}$ consists of two delta-peaks approximately
located at $\pm \frac{\tilde{U}}{2}+ \frac{t_\vec{p}}{2}-\mu$ with
weight $\frac{1}{2}\pm  \frac{t_\vec{p}}{2 \tilde{U}}$.

\begin{figure}
\includegraphics[width=0.9 \linewidth,clip]{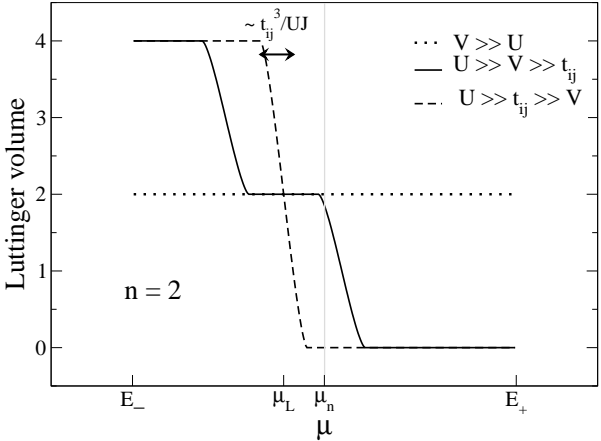}
\caption{Schematic plot of the Luttinger volume [right-hand side
of Eq.~(\ref{luttinger})] as a function of the chemical potential
$\mu$. As $\mu$ is varied  between the lower and upper edge of the
gap, $E_-\le \mu \le E_+$, the density of electrons, $n=2$,
remains fixed. While for a band insulator (dotted line)
Luttinger's theorem ($n$=Luttinger volume) is valid, this is not the
case for strong interactions, where it is either fulfilled only
for a single value of $\mu$, $\mu=\mu_L$, (dashed line) or only a
limited range of chemical potentials (solid line). For fixed $n$
and $T\to 0$, the chemical potential takes the value
$\mu_n=(E_-+E_+)/2$ which coincides with $\mu_L$ only for a
particle-hole symmetric system. \label{figLutVolume}}
\end{figure}

It is easy to see that  Luttinger's theorem is {\em not} valid
for the Mott insulator discussed above. When the chemical
potential is changed {\em within the gap}, $E_-<\mu <E_+$,
the total density of electrons, $n=2$, on the left-hand side of
Eq.~(\ref{luttinger}) remains constant at $T=0$. Therefore the choice
of the chemical potential is completely arbitrary in a grand-canonical
ensemble at $T=0$,
(see Sec.~\ref{mu} for a discussion of a canonical ensemble with fixed
particle number). At the same time, the right-hand side of
Luttinger's sum rule, Eq.~(\ref{luttinger}), changes from $4$ to $0$
and is {\em not} constant, see Fig.~\ref{figLutVolume}. Using for
example the leading order strong-coupling expansion, Eq.
(\ref{leading}), one gets $G(\w=0,\vec{p})>0$ for {\em all}
momenta if $E_-<\mu<0$ and $G(\w=0,\vec{p})<0$ for $0<\mu<E_+$
where
\begin{eqnarray}
E_-&\approx& -\frac{\tilde{U}}{2}+\frac{\max_\vec{k}[t_\vec{k}]}{2}+
O\!\left(\frac{t^2}{J},\frac{t^2}{U}\right) \nonumber \\
E_-&\approx& \frac{\tilde{U}}{2}+\frac{\min_\vec{k}[t_\vec{k}]}{2}+
O\!\left(\frac{t^2}{J},\frac{t^2}{U}\right)
\label{epm}
\end{eqnarray}
are the lower and upper edges of the spectral gap obtained from
Eq.~(\ref{leading}) [due to the finite gap in the system, the
strong coupling expansion is well behaved, see next subsection].

It is well known that for the Hubbard-I approximation, the Fermi
volume of the {\em doped} Hubbard model is not constant in
violation of Luttinger's theorem (see e.g.
Ref.~\cite{jonesMarch}). This is an artifact of the
Hubbard-I approximation. In contrast, we perform a controlled
strong-coupling expansion, which shows (together with simple
general arguments, see also Sec.~\ref{discussion}) that, the
volume within the {\em Luttinger surface} of the undoped system is
not fixed by the particle number.

This proves that
Luttinger's theorem is not valid for a Mott insulator. One may,
however, ask the following questions which will be addressed below:
(i) What happens if higher orders in the strong-coupling expansion are
considered such that the self-energy acquires a momentum dependence?
(ii) What happens in a canonical ensemble where the chemical potential
is fixed by taking the limit $T\to 0$ at fixed particle number $n$? (iii) Which
assumption underlying the proof of Luttinger's theorem is not valid?

\subsection{Finite bandwidth} It is convenient to discuss an
expansion in the hopping $t_{ij}$ using the Greens function in
real space, $G_{ij}(\w)$, where $i$ and $j$ are site indices. To
order $(t_{ij})^2$ only the local Greens function, $G_{ii}$ gets
non-trivial corrections and therefore the self-energy remains
local. The first correction to $G_{ij}$, $i\neq j$, which is not
captured by (\ref{leading}), arises to order $(t_{ij})^3$ when an
electron correlates the sites $i$ and $j$ by three hopping
processes. Close to the poles of $G_{\text{loc}}(\w)$, at $\w+\mu
\sim \pm \tilde{U}/2$, one has to resum perturbation theory in
$t_{ij}$ to infinite order [to obtain e.g. the next correction to
Eq.~\ref{epm})] but this is not necessary in the center
of the gap where the zeros of the Greens function are located.

Simple power-counting is not sufficient to get the precise
dependence of the $(t_{ij})^3$ contribution on $U$ and $J$. In the
limit $t_{ij} \ll J\ll U$, by solving a two-site system
exactly  we obtain $G_{ij}(\w+\mu=0)\approx - 12 (t_{ij})^3/(U^3 J)$. Comparing this
to $G_{\text{loc}}(\w) \approx (\w+\mu)/(\tilde{U}/2)^2$ for small
$\w+\mu$, one finds after a Fourier transformation that the zeros of
$G(\vec{p},\w)$ [and therefore also the poles of $\Sigma(\vec{p},\w)$]
are approximately located at
\begin{align}\label{t3}
\w+\mu = E^*_{\vec{p}} \approx 3 \sum_i \frac{(t_{i0})^3}{J U}
\cos(\vec{p}\cdot \vec{r}_i)+ O\!\left(\frac{(t_{ij})^3}{U^2},\frac{(t_{ij})^4}{U
J^2}\right)
\end{align}
where $\vec{r}_i$ is the vector pointing from site $0$ to site $i$. By
definition, the Luttinger surface is given by $\mu = E^*_{\vec{p}}$
and its volume depends obviously strongly on the value of the chemical
potential $\mu$ in a regime where $n=2=const.$ as is shown in
Fig.~\ref{figLutVolume}. As discussed above, Luttinger's sum rule is
therefore violated.

 Eq.~(\ref{t3}) is consistent
with results of
 Pairault, S\'en\'echal and
Tremblay \cite{tremblay} who performed a strong-coupling expansion
for the {\em single band} Hubbard model [in combination with a
resummation based on a continued-fraction expansion of
$G(\vec{p},\w)$]. In their expression, the $J$ of Eq.~(\ref{t3})
is replaced by the temperature $T$ reflecting the degeneracy of
the strong-coupling ground-state of the single-band Hubbard model.

\subsection{Canonical ensemble and position of the chemical potential}\label{mu}
Within the Mott gap, there is precisely one value of the chemical
potential $\mu=\mu_L$ where Luttinger's sum rule,
Eq.~(\ref{luttinger}) is fulfilled (using again the parameters of
Sec.~\ref{local} in this section). Furthermore, in a canonical ensemble (i.e. for
fixed particle density $n$) the chemical potential has a well defined
limiting value for $T \to 0$, $\mu_n = \lim_{T\to 0} \mu(n,T)$.
Therefore the question arises, whether the Luttinger theorem is valid
for a Mott insulator if a canonical ensemble is used where $\mu=\mu_n$
at $T=0$. Note that in a three-dimensional system, the electron
density is fixed by the long-ranged Coulomb interaction and the
background charge of the ions, therefore the chemical potential does
take the value $\mu_n$ for $T\to 0$. Before showing that the Luttinger
theorem is not valid in this case as generically $\mu_L\neq \mu_n$,
one should first note that in the classical proofs of Luttinger's
theorem \cite{luttinger,dzyaloshinskii} always grand-canonical
ensembles (fixed $
\mu$) and never canonical ensembles
(fixed $n$) are used and therefore there is little reason to believe
that the Luttinger theorem is only valid for a canonical ensemble.

For a particle-hole symmetric situation (often studied in
literature \cite{tsvelik1,tsvelik2,berthod,stanescu}), $\mu_n$ and
$\mu_L$ coincide by symmetry: the symmetry fixes both the value of
$\mu$ and the position and shape of the Luttinger surface completely.
Generically, particle-hole symmetry is not present. To show that
$\mu_n$ and $\mu_L$ differ from each other in this case, we calculate
both of them to leading order in the strong-coupling expansion.

In the canonical ensemble, the chemical potential in located precisely
in the middle of the upper and lower edge of the gap, as the number of particle
excitations $\propto e^{-(E_+-\mu)/T}$ has to be equal to the number
of holes $\propto e^{-(\mu-E_-)/T}$ for $T\to 0$ (see also
Fig.~\ref{figLutVolume}).
Using Eq.~(\ref{epm}) one therefore obtains
\begin{align}
\mu_n=\frac{E_+ + E_-}{2}\approx \frac{\max[
  t_\vec{p}]+\min[t_\vec{p}]}{4} + O\!\left(\frac{t^2}{J},\frac{t^2}{U}\right)
\end{align}
This has to be compared to Eq.~(\ref{t3}) which shows that
$\mu_L=0+O[(t_{ij})^3/(U J)]$. We have therefore proven that $\mu_L
\neq \mu_n$ in the absence of particle-hole symmetry when $\max[
t_\vec{p}]\neq - \min[t_\vec{p}]$. Accordingly, the Luttinger theorem
is violated in the Mott insulating phase even for a canonical ensemble
(in the absence of particle-hole symmetry).

Note that ambiguities in the definition of the chemical potential
and in the frequency, where the Greens function matrix is
evaluated within Luttinger's theorem, are absent in systems where
a Fermi surface exists. Adding to our insulating model
(\ref{Htot}) a further {\em metallic band}, which hybridizes only
weakly with the two bands of the Mott insulating state, does not
change any of our conclusions but fixes $\mu$ unambiguously even
at $T=0$

\subsection{Analysis of the proof} We will now investigate the question, which assumption
underlying the proof of  Luttinger's theorem is not valid for a
Mott insulator. The central step of the
proof \cite{luttinger,dzyaloshinskii} (sketched in
Appendix~B) is to show that the integral
\begin{align}\label{integral}
\int_{-\infty}^{\infty} \frac{d\w}{2 \pi} \, \Sigma(\vec{p},i \w)
\frac{\partial}{\partial i \w} G(\vec{p},i \w)
\end{align}
vanishes. The basic idea is to use that the self-energy can be
written as a derivative of the Luttinger-Ward functional with
respect to the Greens function such that the relevant integrand
can be identified with a total derivative. Note that in
(\ref{integral})
 the integrand drops rapidly with $1/\w^2$ for large frequencies
and that (with the possible exception of the point $\w=0$) no
singularities are expected on the integration path along the
imaginary axis. As Luttinger's theorem is violated in the local
limit, one can use the exact Greens function and self energy for
$t_{ij}=0$, given by Eqs.~(\ref{gLoc}) and (\ref{sigmaLoc}), to
calculate (e.g. by a trivial contour integration)
\begin{align}\label{integralD}
\int_{-\infty}^{\infty} \frac{d\w}{2 \pi} \, \Sigma_{\text{loc}}(i
\w) \frac{\partial}{\partial i \w} G_{\text{loc}}(i \w)
=-\frac{1}{2} \, \text{sign} \, \mu
\end{align}
This well-behaved integral does {\em not} vanish, contrary to what
is suggested by the proof of Luttinger's theorem. Taking into
account  spin- and orbital summations, the factor $\pm 1/2$
explains the violation of  Luttinger's theorem by $\pm 2$ for
$\mu\neq 0$ in the local limit.

\begin{figure}
\includegraphics[width=0.95 \linewidth,clip]{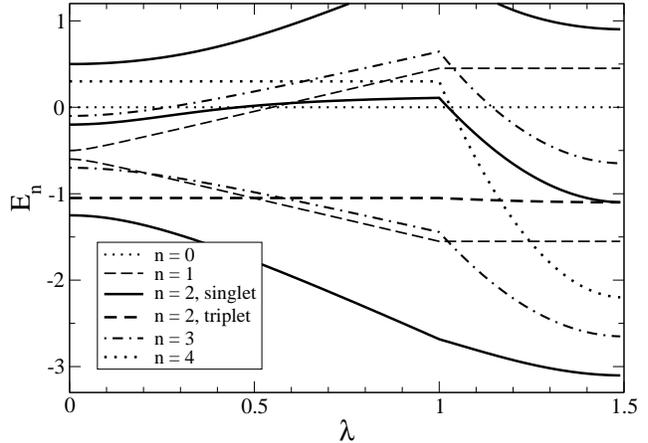}
\caption{Evolution of the energy levels in the local limit,
$t^\alpha_{ij}=0$, see Appendix~A, as a function of
a control parameter $\lambda$. For  $\lambda=0$, the inter-orbital
hybridization $V$ vanishes, $V=0$, and $U_1=1, U_2=1.5, \mu=0,
J=0.2, \epsilon_1=-0.6, \e_2=-0.5$. For $0\le \lambda \le 1$,
$V=\lambda$ increases linearly with $\lambda$ and the interactions and
$\mu$
stay constant. For $1 \le \lambda \le 1.5$, $V=1$ is constant
\cite{footnote} but the interactions $U_1$, $U_2$ and $J$ drop
proportional to $4 (1.5-\lambda)^2$. For $\lambda=1.5$, the
non-interacting system is characterized by two single-particle
levels with energies $\frac{\epsilon_1+\epsilon_2}{2} \mp
\sqrt{\left(\frac{\epsilon_1-\epsilon_2}{2}\right)^2+V^2}\approx
-1.55$ and $0.45$. The groundstate is always a singlet with two
electrons per site, $n=2$
well separated by a finite gap from all
excited state. As there is no level crossing, 
the wave function evolves smoothly from a singlet made from
localized electrons in two different bands (Mott insulator for
small $\lambda$) 
to two electrons filling a single band (band insulator at $\lambda=1.5$).
The finite gap to excited states guarantees that also in the presence
of small but finite hopping
 $t^\alpha_{ij}$ two-orbital Mott- and band-insulator are
 adiabatically connected along the described path. \label{figLevels}}
\end{figure}
It is presently not  clear, where precisely the problem is located
in the arguments  which try to show that (\ref{integral}) vanishes
(see Appendix B). Despite the simple form of the self
energy (\ref{sigmaLoc}), bare
 perturbation theory is very singular for $\mu\neq 0$, $T=0$, as
the density of the half-filled model, $n=2$, is reached only above
a critical value of $U$. Inspection of the exact finite
temperature Greens function in the local limit indeed shows
divergencies with powers of $1/T$ to arbitrary order. One may
therefore suspect that the replacement of Matsubara sums by
integrals, i.e. the limit $T\to 0$, is problematic. Similarly, it
may not be sufficient to show that (\ref{integral}) vanishes order
by order in a skeleton expansion (see Appendix B) for
such a non-perturbative problem. It is, however, interesting to
remark, that the Luttinger-Ward functional can in principle be
constructed non-perturbatively as pointed out by
Potthoff \cite{pothoff} who also emphasized in his paper that the
proof of  Luttinger's theorem requires non-trivial assumptions on
the regularity of the limit $T \to 0$. A more detailed discussion
of assumptions underlying the proof can be found in Appendix~B.

In Ref.~\cite{altshuler}, Altshuler {\it et al.} studied
the question, to what extent Luttinger's theorem applies to an
antiferromagnetic metal if one insists to describe the system {\em
without} explicitly breaking the symmetry. The authors found that
in this situation the integral (\ref{integral}) does not vanish.
In their case, the problem could be traced back to an anomaly
(i.e. a contribution arising after the proper regularization of a
divergence) which seems not to be the case for a Mott insulator
with unique ground state studied here.

\section{Adiabatic continuity}
Based on the results discussed above, we will now investigate the
crossover from a band- to a two-orbital Mott insulator (or Kondo
insulator) with two electrons per unit cell, $n=2$. Note that we
are {\em not} discussing the standard single-orbital Mott
insulator with an odd number of electrons per unit cell, which in
$d>1$ usually becomes magnetically ordered for $T=0$. For a band
insulator Luttinger's theorem is always trivially fulfilled and
there is no Luttinger surface for any value of $\mu$. In contrast,
we have shown that for $V,t_{ij} \ll U, J$, Luttinger's theorem is
violated for almost all values of $\mu$ within the gap. How does
the crossover happen or is there a quantum phase transition
between these two qualitatively different states? This question is
answered by increasing $\sqrt{V^2+(\e_1-\e_2)^2}$ in the model
(\ref{Htot}) in the trivial limit of small $t_{ij}$.

We first investigate the evolution of the ground state when
changing the parameters smoothly starting from two coupled Mott
insulators ($V=0$, $U_1=1, U_2=1.5, J=0.2$) and ending with a band
insulator ($V=1$, $U_i=0, J=0$). For $t^\alpha_{ij}=0$, it is
shown in Fig.~\ref{figLevels} that on this adiabatic path the
ground state remains a singlet always separated by a finite gap
$\Delta$ from all excited states (as guaranteed by level
repulsion). As the gap is finite, these conclusions hold also in
the presence of a small but finite band-width, $t^\alpha_{ij} \neq
0$. This proves that there is no quantum-phase
transition \cite{footnote}, separating the two-orbital Mott
insulator from the band insulator, instead there is just a smooth
crossover (consistent with results in the literature, see e.g.
Ref.~\cite{monien2}).

As an aside, we also note that it is generally believed that the
magnetically ordered states of the particle-hole symmetric,
half-filled one-band Hubbard model at small $U$ (spin-density wave
insulator) and large $U$ (magnetically ordered Mott insulator) are
adiabatically connected.

\begin{figure}
\includegraphics[width=0.9 \linewidth,clip]{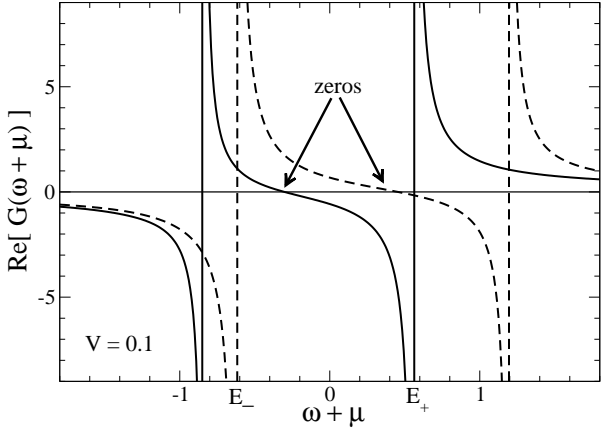}
\caption{Both eigenvalues of the Greens-function matrix (solid and
  dashed) as a function of $\w+\mu$ in the local limit
  $t^\alpha_{ij}=0$ for $V=0.1$, $U_1=1, U_2=1.5, J=0.2,
  \epsilon_1=-0.6, \e_2=-0.5$ and a chemical potential within the gap,
  $E_-<\mu<E_+$. Both Greens functions change their sign within the
  gap for this value of $V$.  Luttinger's theorem holds only if the
  chemical potential is located between those zeros, see middle panel
  of Fig.~\ref{figZeros}. \label{figGreen}}
\end{figure}

To investigate the role of Luttinger's theorem and the Luttinger surface
as a function of the band splitting given by $\sqrt{V^2+(\e_2-\e_1)^2}$, we
first consider the effect of a potential $\Delta \e=\e_2-\e_1$ which
shifts the two bands relative to each other, $\e_{1/2}=-U/2\pm \Delta
\e/2$, in the limit $V= t^\alpha_{ij}=0$ and $U_1=U_2=U$. The total
charge gap is given by $U-\Delta \e$ and the zeros of the Greens
functions of the two orbitals split by $\Delta \e$. For $\Delta \e <
U/2$, the two zeros remain within the gap. While
Luttinger's theorem is valid for $-\Delta \e/2<\mu < \Delta \e/2$, it
is violated outside of this regime as is shown in
Fig.~\ref{figLutVolume} (for $t_{ij}=0$ the two jumps in the solid
line become sharp step functions).  For $U/2<\Delta \e<U$, Luttinger's
sum rule~(\ref{luttinger}) is valid for all values of $\mu$ within the
gap but the system is still a Mott insulator in the sense that each of
the two bands is singly occupied. For
$\Delta \e=U$, two levels cross and the ground-state wave-function
changes in a first order transition which is an artifact of the limit
$V=0$.  Therefore this limit does not
describe a smooth transition from a two-orbital Mott- to a band
insulator, which can, however, be obtained by increasing the
inter-orbital hybridization $V$.

\begin{figure}
\includegraphics[width=0.9 \linewidth,clip]{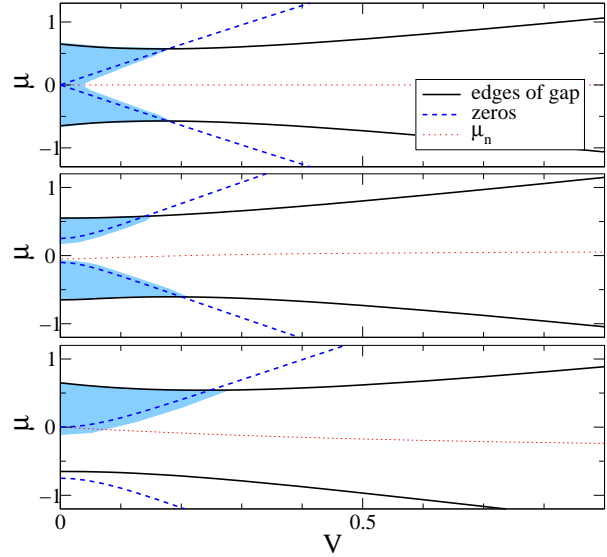}
\caption{ Crossover from the Mott insulator (small
intra-orbital hybridization $V$) to a band insulator (large $V$).
For a Mott insulator with a chemical potential in the shaded
region, Luttinger's theorem is violated (with the exception of
the line $\mu=0$ in the upper panel). Luttinger's theorem is, however, valid for a band
insulator at large $V$. Solid lines denote the upper and lower
band edge, the dashed lines the positions of the poles of the
self-energy or equivalently of the zeros of the two Greens
functions (see Fig.~\ref{figGreen}) obtained from the exact
solution of the local model, $t^\alpha_{ij}=0$, sketched in
Appendix~A. In the plot the shaded region is not
exactly delimited by the dashed lines to take schematically a
small but finite hopping $t^\alpha_{ij}$ into account (c.f.
Fig.~\ref{figLutVolume}). The dotted lines denote the position of the
chemical potential for fixed particle number, $\mu_n=\lim_{T\to 0}
\mu(n,T)$. Upper panel: particle-hole symmetric case
with $U_1=U_2=1,
\epsilon_1=\epsilon_2=-1/2, J=0.2, t^\alpha_{ij}=0$. Middle panel:
same parameters as in Fig.~\ref{figLevels}. Lower panel:
$U_1=1, U_2=2.5, \e_1=-0.5, \e_2=-2, J=0.2,  t^\alpha_{ij}=0$. \label{figZeros}}
\end{figure}
Increasing $V$ has a very similar effect as increasing $\Delta \e$
for $\Delta \e < U$ as is shown in Figs.~\ref{figGreen} and \ref{figZeros}. The shaded
regions in  Fig.~\ref{figZeros} indicate the range of chemical
potentials where Luttinger's theorem does not hold. As discussed
above, also a finite but small hopping $t^\alpha_{ij}$ does not
change the picture qualitatively  (see Figs.~\ref{figLutVolume}
and \ref{figZeros}).

\section{Discussion}\label{discussion}

The main message of this paper is negative: Luttinger's theorem
in the form given by Eq.~(\ref{luttinger}) is not valid in a Mott
insulator and therefore the concept of a Luttinger surface (points
in momentum space where the Greens function vanishes at $\w=0$)
seems not to be very useful --- especially when compared to the
much more powerful concept of a Fermi surface.

The basic argument is that  a change of the chemical potential
within the charge gap, $E_-<\mu<E_+$, just shifts the frequency
argument of all Greens functions
\begin{eqnarray} G(\w,\vec{p},\mu)=G(\w+\mu,\vec{p})
\end{eqnarray}
If the Greens functions change their sign within the gap (as is
the case for a Mott insulator), this implies that the sign of
$G(\w=0,\vec{p})$ and therefore the Luttinger volume depends on
the arbitrary choice of $\mu$ while the density of electrons
remains constant. We have also shown, that even in the canonical
ensemble, when the chemical potential is fixed to $\lim_{T\to
  0}\mu(n,T)$ by the particle density $n$, the Luttinger theorem is
violated in the absence of particle-hole symmetry.

Often Luttinger's theorem is stated in a different way: the
volume enclosed by the {\em Fermi surface} (defined by poles
rather than sign changes of the Greens functions) is given by the
number
 of electrons per unit cell modulo $2$ (to take into account filled
 bands). For this case alternative proofs
 exist \cite{oshikawa,praz} which are based on Gauge invariance,
 topological arguments and the assumption that a
 Fermi liquid exists \cite{oshikawa} or on particle-number conservation
 and an adiabatic connection of the interacting to the non-interacting
metal \cite{praz}.
 As we have consided only insulators with $n=2$, Luttinger's
 theorem in this  ``Fermi-surface version'' is trivially valid.
Recently, Senthil, Sachdev and Vojta \cite{senthil} argued that one
can construct a situation where this is not the case. They
considered  a two-band model, where the first band is in its Mott
insulating state such that it can be described by free spins.
Senthil, Sachdev and Vojta then assumed that one can add strongly
frustrating interactions between those spins such that the spins
form a {\em gapped} spin liquid without breaking any symmetry. If
the second band is weakly interacting, it will form a Fermi
surface. As the spin-liquid is gapped, this state is robust
against small perturbations like a finite coupling to the Fermi
liquid of the second band, e.g. by a small $V$. In such a situation, the volume
enclosed by the Fermi surface is given by the number of electrons
in the second band $n_2$ rather than by the total number of
electrons per unit cell, $n=1+n_2$. Combining this argument with
the results of our paper, one realizes that in such a case both
versions of  Luttinger's theorem are not valid (in the absence of
particle-hole symmetry).

Luttinger's theorem can also break down, if there is a finite
imaginary part of the self energy at $\w=0$. A large imaginary
part has for example been observed in numerous studies of weakly
doped Mott insulators at small, but finite temperatures, see e.g.
Refs.~\cite{haule}, or in almost magnetic systems, e.g.
Ref.~\cite{vlik}. Also in a slightly doped tJ model a violation of the
Luttinger theorem has been observed in small systems \cite{prelovsek}.
It may be worthwhile to emphasize that
the breakdown of Luttinger's theorem discussed here is of
 different origin and probably unrelated.

Our analysis of Luttinger's theorem, Eq.~(\ref{luttinger}),
appears to imply a qualitative difference between a band-insulator
where it is  valid and a two-orbital Mott insulator where it
breaks down. Based on the ``phase diagrams'' of
Fig.~\ref{figZeros} one could define a ``critical'' hybridization
$V_c$, e.g. by demanding that Luttinger's theorem is valid for all
$V \ge V_c$. We have, however, shown that at this ``critical''
hybridization, the ground-state energy (and essentially all
observables with the exception of the Luttinger surface, the
Luttinger volume and integral (\ref{integral})) is analytic in $V-V_c$: no phase transition
takes place. Instead there is a smooth crossover from a
two-orbital ($n=2$) Mott- to a band-insulator. Recently, Konik,
Rice and Tsvelik \cite{tsvelik2} investigated coupled Mott ladders
 constructing explicitly a state where simultaneously a Luttinger
surface and several Fermi surfaces (due to small electron and hole
pockets) are present. While we have not studied such a situation,
in analogy to the results presented here, we speculate that the
system studied in Ref.~\cite{tsvelik2} is adiabatically
connected to some non-interacting model where band-structure
effects lead to the formation of particle and hole pockets.

While we have shown in this paper that band- and two-orbital Mott
insulators can in principle be adiabatically connected, one should
keep in mind that for many models there will be a series of first
and/or second-order phase transitions to various metallic, insulating
and/or symmetry broken states when the interactions are increased. The
best studied case in this context is probably the ionic Hubbard model,
see Refs. \cite{ionic,ionic1,ionic2,ionic3} and references
therein. Note, however, that the symmetries of this special model
ensure \cite{ionic}, that the Mott insulating phase has no spin
gap in contrast to the case investigated in this paper.

For the future, it would be interesting to nail down more
precisely under what conditions  Luttinger's theorem breaks down
and which of the assumptions used in various proofs is most
fragile. This is especially important, as Luttinger's theorem
can be a powerful tool to classify topologically different states
of matter at least for metallic systems.

\begin{acknowledgement}
 I would like to thank A. Altland, I. Fischer, R.~Helmes,
 E. M\"uller-Hartmann,  A. A. Nersesyan, N.~Shah, A.~M.~Tsvelik,
 M. Vojta, and J. Zaanen for useful
discussions and the SFB 608 of the DFG for financial support.
\end{acknowledgement}

\appendix
  \renewcommand{\theequation}{A.\arabic{equation}}
  \setcounter{equation}{0}  
\section*{Appendix A: Spectrum of $H_{\text{loc}}^i$}\label{AppLevels}

The spectrum of the local Hamiltonian is given by the following
energies: The empty site, $n=0$, has vanishing energy, $E_0=0$,
while one obtains $E_4=2 (\e_1+\e_2)+U_1+U_2 -4 \mu$ for $n=4$.
The energies of the four single-particle states, $n=1$, are $E_{1
\pm}=\left(\frac{\epsilon_1+\epsilon_2}{2} \right)\pm
\sqrt{\left(\frac{\epsilon_1-\epsilon_2}{2}\right)^2+V^2} - \mu$,
while the 4 three-particle states have  $E_{3
\pm}=\e_1+\e_2+\left(\frac{\epsilon_1+\epsilon_2+U_1+U_2}{2}
\right)\pm
\sqrt{\left(\frac{\epsilon_1+U_1-\epsilon_2-U_2}{2}\right)^2+V^2}
- 3 \mu$. The energy of the three $n=2$ triplet states is given by
$E_{2t}=\e_1+\e_2+\frac{J}{4}-2 \mu$. Furthermore there are three
singlet  states for $n=2$ with energies $E_{2s1}<E_{2s2}<E_{2s3}$
which can be determined by diagonalizing the $3\times 3$ matrix
\begin{equation} \left(\begin{array}{ccc}
2 (\e_1-\mu)+U_1 &0 & \sqrt{2} V \\
0 & 2 (\e_2-\mu)+U_2 &  \sqrt{2} V \\
 \sqrt{2} V &  \sqrt{2} V & \e_1+\e_2-2 \mu -3 J/4
  \end{array}\right) \end{equation}
The analytic formulas
for $E_{2si}$ can be obtained by solving a cubic equation. As they are
not very instructive and rather long, they are not displayed here. We only note that level
repulsion ensures that the three
energies are not degenerate for $V\neq 0$.

The total size of the gap for charge excitations of the local model with
ground-state $n=2$ is determined by $E_{1-}+E_{3-}-2 E_{2s1}$. For $V=0$ it
is for example given by  $\frac{3}{2} J+ \min[\e_1+U_1,\e_2+U_2]
-\max[\e_1,\e_2]$.

From the eigenfunctions and eigenvalues of the Hamiltonian, one
can easily obtain the Greens function.

  \renewcommand{\theequation}{B.\arabic{equation}}
  \setcounter{equation}{0}  
\section*{Appendix B: A proof of  Luttinger's theorem}\label{proof}
In this appendix we briefly sketch a version of the ``proof'' of
Luttinger's theorem. The first step is to insert
$1=\frac{\partial}{\partial i \w} (i \w-\epsilon_\vec{k})$ in the
expression for the density of electrons using imaginary
frequencies \cite{dzyaloshinskii} at $T=0$
\begin{eqnarray}
n&=& \lim_{T\to 0, V\to \infty} \frac{2 T}{V} \sum_{\vec{p}}
\sum_{\w_n} e^{i \w_n \epsilon} G(\vec{p},i
\w_n)\\
&=&   \int_{\w\vec{p}} \frac{2e^{i \w \epsilon}}{(2
\pi)^{d+1}}G(\vec{p},i \w) \frac{\partial}{\partial i \w} \left[
G^{-1}(\vec{p},i
\w)+\Sigma(\vec{p},i \w) \right] \nonumber\\
&=&  \frac{2}{(2 \pi)^{d+1}} \int_{\w\vec{p}} e^{i \w \epsilon}
\frac{\partial}{\partial i \w} \ln[G^{-1}(\vec{p},i \w)] \label{nn}\\
&& - \frac{2}{(2 \pi)^{d+1}} \int_{\w\vec{p}} \int_\w
\Sigma(\vec{p},i \w) \nonumber \frac{\partial}{\partial i \w}
G(\vec{p},i \w)
\end{eqnarray}
where $\int_{\w\vec{p}}=\int_{-\infty}^\infty d\w
\int_{1^{st}\text{ BZ}} d^d \vec{p}$ describes the integration of
frequencies along the imagninary axis and of momenta over the
first Brillouin zone. For the last equation, we have performed a
partial integration using that $\lim_{\w \to \pm \infty}
\Sigma(\vec{p},i \w) G(\vec{p},i \w) =0$. Using the usual analytic
continuation arguments \cite{luttinger,dzyaloshinskii} and the
vanishing of $\text{Im} \Sigma(\vec{p},\w=0)$ one can identify the
first term in (\ref{nn}) with the right-hand side of
Eq.~(\ref{luttinger}) [to keep the presentation simple, we omit
the discussion of band indices \cite{luttinger,dzyaloshinskii}].
Luttinger's theorem is therefore valid if the second term in
(\ref{nn}) vanishes -- which we have shown to be not the case for
a generic Mott insulator.

Let us nevertheless try to ``prove'' it. The self energy can be
obtained by varying the Luttinger-Ward functional \cite{luttinger}
$\Phi[G]$, $\Sigma(\vec{p},i \w)=\frac{\delta \Phi}{\delta
G(\vec{p},i \w)}$. The Luttinger-Ward functional is given by the
sum of all skeleton diagrams. By the equation
$\Phi[G]=\int_{\w\vec{p}}G(\vec{p},i \w) {\Phi}_G(\vec{p},i \w) $
we define the functional ${\Phi}_G$ with
\begin{eqnarray}
\Sigma(\vec{p},i \w)={\Phi}_G(\vec{p},i \w)+\int_{\w'\vec{p}'}
\!\! G(\vec{p}',i \w') \frac{\delta {\Phi}_G(\vec{p}',i
\w')}{\delta G(\vec{p},i \w)} \label{B3}
\end{eqnarray}
Consider the following integral over a total derivative
\begin{eqnarray}\label{B4}
0&=&\int_{\w\vec{p}} \frac{d}{d i \w}\left[ G(\vec{p},i \w)
{\Phi}_G(\vec{p},i \w) \right]
\end{eqnarray}
By inspecting the self-energy style diagrams contributing to
${\Phi}_G(\vec{p},i \w)$ one realizes that a change of the
external frequency $\w$ can be absorbed by a change of the
frequency of all internal lines and therefore
\begin{align}
\frac{d}{d i \w} {\Phi}_G(\vec{p},i
\w)=\int_{\w'\vec{p}'}\frac{\delta {\Phi}_G(\vec{p},i \w)}{\delta
G(\vec{p}',i \w')} \frac{d}{d i \w'} G(\vec{p}',i \w')
\end{align}
One obtains
\begin{eqnarray} 0&=&\int_{\w\vec{p}}
\frac{d}{d i \w}\left[ G(\vec{p},i \w) \nonumber
{\Phi}_G(\vec{p},i \w) \right]\\
&=&\int_{\w\vec{p}}\Bigl[ {\Phi}_G \frac{d}{d i \w} G  \nonumber \\
&&+G \int_{\w'\vec{p}'}\frac{\delta {\Phi}_G(\vec{p},i \w)}{\delta
G(\vec{p}',i \w')} \frac{d}{d i \w'} G(\vec{p}',i \w')\Bigr]
\nonumber
\\
&=& \int_{\w\vec{p}}\Sigma(\vec{p},i \w) \frac{\partial}{\partial
i \w} G(\vec{p},i \w) \label{B6}
\end{eqnarray}
where in the last step we have exchanged the $\w, \vec{p}$ with
the $\w', \vec{p}'$ integration, renamed the variables and used
Eq.~(\ref{B3}). If all arguments leading to (\ref{B6}) are valid,
then Luttinger's theorem is proven by Eq.~(\ref{nn}).

However, in the main text, we have shown by an explicit example
that (\ref{B6}) does not vanish. Therefore one of the steps in
(\ref{B6}) is not valid for a Mott insulator. The first question
is whether all integrals are well defined. This seems to be the
case as for $\mu\neq 0$ there is no singularity on the imaginary
axis and at least order by order in the skeleton expansion each
term vanishes with $1/\w^2$ for $\w \to \pm \infty$. Also
exchanging the order of integration is probably valid order by
order in the skeleton expansion, as all integrals seem to be
absolutely converget. As far as we can see, only two problems
remain. Either the skeleton expansion used frequently above is
highly singular or the problem is hidden in the limit $T\to 0$ and
the use of integrals (instead of Matsubara sums) on the frequency
axis and the use of derivatives with respect to frequencies was
not correct. In the main text, it is  argued that bare
perturbation theory is very singular for $T\to 0$, indicating that
the two problems are related.



\begin{thebibliography}{}
\bibitem{footnote} The non-analytic kinks in
  Fig.~\ref{figLevels} are a trivial artifact of the non-analytic
  dependence of the bare coupling constants on $\lambda$. They do
  therefore not signal a quantum phase transition.
\end{thebibliography}
\end{document}